\documentclass[journal]{IEEEtran}

\usepackage{amsmath}
\usepackage{amssymb}
\usepackage{graphicx}
\usepackage{cite}
\usepackage{siunitx}
\usepackage{tumcolor}
\usepackage[nolist]{acronym}

\usepackage{amsmath}
\usepackage{amsthm}
\usepackage{amssymb}
\usepackage{bm}
\usepackage{xspace}
\usepackage{xcolor}
\usepackage{graphicx}
\usepackage{dsfont}

\usepackage{pgf}
\usepackage{pgfplots}
\usepackage{pgfplotstable}

\usetikzlibrary{arrows,shapes.misc,chains,scopes,intersections}
\usetikzlibrary{matrix}
\usetikzlibrary{calc}
\usetikzlibrary{fit}
\usetikzlibrary{graphs}
\usetikzlibrary{shadows.blur}
\usetikzlibrary{shapes.symbols}
\usepgfplotslibrary{patchplots}

\usepackage{booktabs}
\usepackage{colortbl}

\graphicspath{{figures/}}

\theoremstyle{plain}

\newcounter{algocount}
\newcounter{examplecount}

\newcommand{\vecB}{\boldsymbol{B}}

\newcommand{\matp}{\boldsymbol{P}}
\newcommand{\mati}{\boldsymbol{I}}

\newcommand{\seta}{\ensuremath{\mathcal{A}}\xspace}

\newcommand{\setx}{\ensuremath{\mathcal{X}}\xspace}

\newcommand{\bmm}{\begin{matrix}}
\newcommand{\emm}{\end{matrix}}
\newcommand{\bpm}{\begin{pmatrix}}
\newcommand{\epm}{\end{pmatrix}}

\newcommand{\bsbm}{\left[\begin{smallmatrix}}
\newcommand{\esbm}{\end{smallmatrix}\right]}
\newcommand{\bspm}{\left(\begin{smallmatrix}}
\newcommand{\espm}{\end{smallmatrix}\right)}

\newcommand{\bbm}{\begin{bmatrix}}
\newcommand{\ebm}{\end{bmatrix}}

\DeclareMathOperator{\expop}{\mathbb{E}}
\DeclareMathOperator{\entop}{\mathbb{H}}
\DeclareMathOperator{\miop}{\mathbb{I}}

\DeclareMathOperator*{\minimize}{minimize}

\DeclareMathOperator*{\st}{subject\;to}

\DeclareMathOperator{\dm}{dm}

\newtheorem{remark}{Remark}

 \newcommand{\snr}{\ensuremath{\mathsf{SNR}}\xspace}

\graphicspath{{figures/}}

\newcommand{\rbmd}{R_\textnormal{bmd}}

\DeclareSIUnit\dBm{dBm}
\DeclareSIUnit{\bpcu}{bpcu}

\newcommand{\rdm}{R_\text{dm}}
\newcommand{\rloss}{R_\text{loss}}
\newcommand{\rtx}{R_\text{t}}

\newcommand{\prbmd}{\rbmd^{\Pi}}
\newcommand{\rbicm}{R_\textnormal{bicm}}

\pgfplotsset{compat=newest}

\newcommand{\tikzsetnextfilename}[1]{}

\begin{document}

\begin{acronym}[AAAAAAA]
\acro{OFDM}{orthogonal frequency division multiplexing}
\acro{DMT}{discrete multitone}
\acro{PDM}{product distribution matching}
\acro{SVD}{singular value decomposition}
\acro{ADSL}{asymmetric digital subscriber line}
\acro{SNR}{signal-to-noise ratio}
\acro{DM}{distribution matcher}
\acro{QAM}{quadrature amplitude modulation}
\acro{SE}{spectral efficiency}
\acro{CCDM}{constant composition distribution matcher}
\acro{BP}{belief propagation}
\acro{LDPC}{low-density parity-check}
\acro{FER}{frame error rate}
\acro{FEC}{forward error correction}
\acro{PS}{probabilistic shaping}
\acro{GS}{geometric shaping}
\acro{PAS}{probabilistic amplitude shaping}
\acro{ATSC}{Advanced Television Systems Committee}
\acro{NUC}{Non-Uniform Constellations}
\acro{MB}{Maxwell-Boltzmann}
\acro{ASK}{amplitude shift keying}
\acro{AWGN}{additive white Gaussian noise}
\acro{NBBC}{natural based binary code}
\acro{NBC}{natural binary code}
\acro{BRGC}{binary reflected Gray code}
\acro{DMS}{discrete memoryless source}
\end{acronym}


\title{High Throughput Probabilistic Shaping\\with Product Distribution Matching}

\author{Georg B\"ocherer,~\IEEEmembership{Member, IEEE}, Fabian Steiner,~\IEEEmembership{Student Member, IEEE}, Patrick Schulte,~\IEEEmembership{Student Member, IEEE}
\thanks{G. B\"ocherer,  F. Steiner, P. Schulte are with the Institute for Communications Engineering, Technical University of Munich (TUM). }
\thanks{The work was partly supported by the German Federal Ministry of Education and Research in the framework of an Alexander von Humboldt Professorship.
}
\thanks{The ideas presented in this work have been filed as a patent application with the EPO (application number: EP16192404.8) on October 5, 2016.}
}

\markboth{}{}%

\maketitle

\begin{abstract}
Product distribution matching (PDM) is proposed to generate target distributions over large alphabets by combining the output of several parallel distribution matchers (DMs) with smaller output alphabets. The parallel architecture of PDM enables low-complexity and high-throughput implementation. PDM is used as a shaping device for probabilistic amplitude shaping (PAS). For 64-ASK and a spectral efficiency of \num{4.5} bits per channel use (bpcu), PDM is as power efficient as a single full-fledged DM. It is shown how PDM enables PAS for parallel channels present in multi-carrier systems like digital subscriber line (DSL) and orthogonal frequency-division multiplexing (OFDM). The key feature is that PDM shares the DMs for lower bit-levels among different sub-carriers, which improves the power efficiency significantly. A representative parallel channel example shows that PAS with PDM is \SI{0.93}{dB} more power efficient than conventional uniform signaling and PDM is \SI{0.35}{dB} more power efficient than individual per channel DMs.
\end{abstract}


\section{Introduction}
\label{sec:intro}

\IEEEPARstart{H}{}igher-order modulation is indispensable in mobile, satellite, cable, and fiber-optic communication to achieve the high \ac{SE} required for data applications.

Transceivers must be flexible, i.e., they should support different \ac{SE}s so they can adapt to the link quality at hand and deliver the best possible connectivity. Conventional coded modulation uses uniform distributions on the constellation points. This has two disadvantages. First, uniform distributions suffer a power inefficiency of up to \SI{1.53}{dB}. Second, flexibility can be achieved only by supporting a large number of modcods, i.e., combinations of modulation formats and channel codes. For example DVB-S2X requires supporting 116 modcods~\cite{etsi2014dvb}.

\begin{figure}[t!]
    \centering
    \tikzsetnextfilename{systemModel_matrix}
	\includegraphics{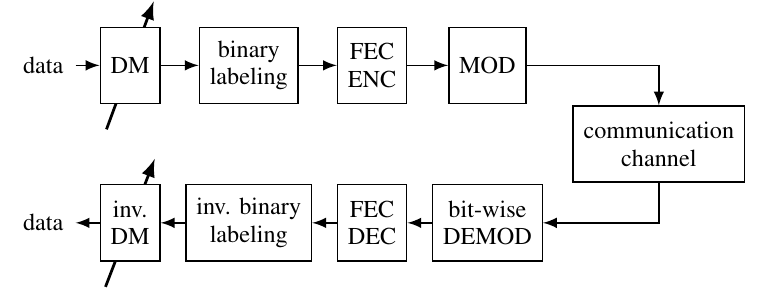}
	\caption{System model of PAS. The shaping device DM is concatenated in reverse with the FEC device.}
	\label{fig:system_model}
    \vspace{0.3cm}
    \tikzsetnextfilename{parallelBinaryMatching2}
    \includegraphics{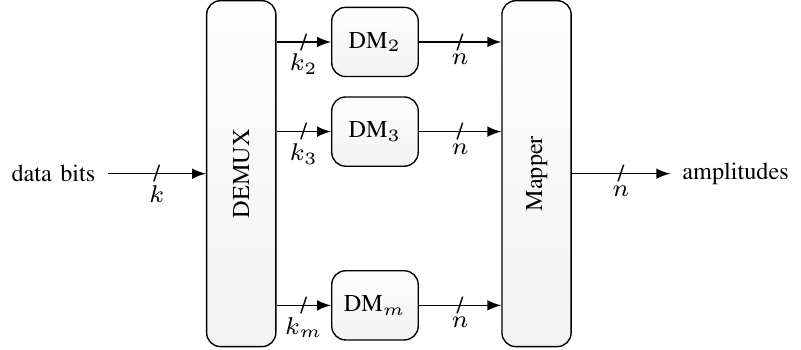}
    \caption{The DM implementation proposed in this work: Product Distribution Matching (PDM) for $2^m$-ASK. $k$ binary data bits are demultiplexed into $m-1$ parallel blocks of sizes $k_2$ to $k_m$. Parallel binary component DMs output $m$ shaped sequences of length $n$. A bit-mapper recombines the $m-1$ sequences and outputs one shaped amplitude sequence of length $n$.}
    \label{fig:pdm}
\end{figure}

One approach that has been proposed is \ac{GS}~\cite{barsoum_constellation_2007,loghin_non-uniform_2016} which uses constellations with non-equidistant signal points. While improved power efficiency was observed, the problem of flexibility remains. A second approach is \ac{PS} that uses equidistant signal points with a non-uniform distribution. For an overview of \ac{PS} schemes, see \cite[Sec.~II]{bocherer2015bandwidth} and references therein. Recently, we proposed \ac{PAS}~\cite{bocherer2015bandwidth}, a \ac{PS} architecture that concatenates a \ac{DM}~\cite{bocherer2011matching,schulte2016constant} as a shaping device with \ac{FEC}, see Fig.~\ref{fig:system_model}. \ac{PAS} achieves the optimal power efficiency and enables flexible \ac{SE} with only one \ac{FEC} code~\cite[Sec.~VIII]{bocherer2015bandwidth}. \ac{PAS} has been successfully integrated with \ac{LDPC} codes~\cite{bocherer2015bandwidth}, turbo codes~\cite{yuan2015rate}, SC-LDPC codes~\cite{buchali2016rate}, polar codes~\cite{prinz2016polar}, and nonbinary codes~\cite{boutros2017probabilistic}. In comparison~\cite{steiner_geom}, \ac{PAS} is over \SI{0.3}{dB} more power efficient than \ac{NUC}~\cite{loghin_non-uniform_2016}, a \ac{GS} implementation advocated by the \ac{ATSC}~3.0 standard. \ac{PAS} is being considered for inclusion in the 5G standard~\cite{huawei_tdoc}. The benefits of \ac{PAS} for fiber-optic communication were recently showcased in a field trial~\cite{idler2017field} and future optical modems will implement \ac{PS}~\cite[Sec. V-A]{roberts_beyond_2017}.

The enabling technology for \ac{PAS} is the \ac{DM}, which transforms a binary data sequence into a sequence of symbols with a desired distribution. For an overview of existing \ac{DM} algorithms, see \cite[Sec.~I]{schulte2016constant} and references therein. For implementation, fixed-to-fixed length \ac{DM}s are desirable. For high-throughput applications, efficient DM encoding is required. Furthermore, fixed-to-fixed length \ac{DM}s require a large block length to work well~\cite{schulte2017divergence}.

In many practical settings, the data link is well modelled by a set of non-interacting parallel channels. Examples include multi-carrier transmission such as \ac{OFDM}, \ac{DMT}, and multi-antenna transceivers when the \ac{SVD} of the channel matrix is used to orthogonalize the system. Employing current \ac{DM} algorithms in such scenarios is challenging, as techniques like bit-loading partition the transmitted sequence in several short segments, each with an individual constellation size and distribution, which potentially causes a significant rate loss.

In this work, we propose a novel \ac{DM} architecture called \ac{PDM}, which internally uses a collection of parallel \ac{DM}s with smaller output alphabets to synthesize the desired distribution as a product distribution. A preferable implementation uses binary output alphabets for the individual \ac{DM}s. This approach both facilitates high-throughput applications by parallelization and reduces the rate loss for short output lengths, which makes the \ac{PDM} particularly amenable for large constellations and high-throughput. In the final part of this work, we propose extended \ac{PDM} for parallel channels, which shares the component \ac{DM}s for lower bit-levels among different sub-carriers. Extended \ac{PDM} can be applied, e.g., in \ac{OFDM} and \ac{DMT}. We provide a representative example where extended \ac{PDM} is \SI{0.93}{dB} and \SI{0.35}{dB} more power efficient than uniform signaling and individual per sub-carrier \ac{DM}s, respectively, and operates close to the waterfilling limit. All simulation results were obtained using the \ac{DM} implementations by \cite{shapecomm}.

This work is structured as follows. Sec.~\ref{sec:prelim} reviews \ac{DM}s and \ac{PAS} and states achievable rate expressions for system design. In Sec.~\ref{sec:pdm}, we introduce the \ac{PDM}
architecture and present finite length simulation results for 64-QAM. Sec.~\ref{sec:parallel_channel} shows how extended \ac{PDM} can be used to operate \ac{PAS} close to the waterfilling limit of parallel
channels. We conclude in Sec.~\ref{sec:conclusion}.

\section{Preliminaries}
\label{sec:prelim}

\begin{figure*}
\centering
\footnotesize
\tikzsetnextfilename{pasx2}
\includegraphics{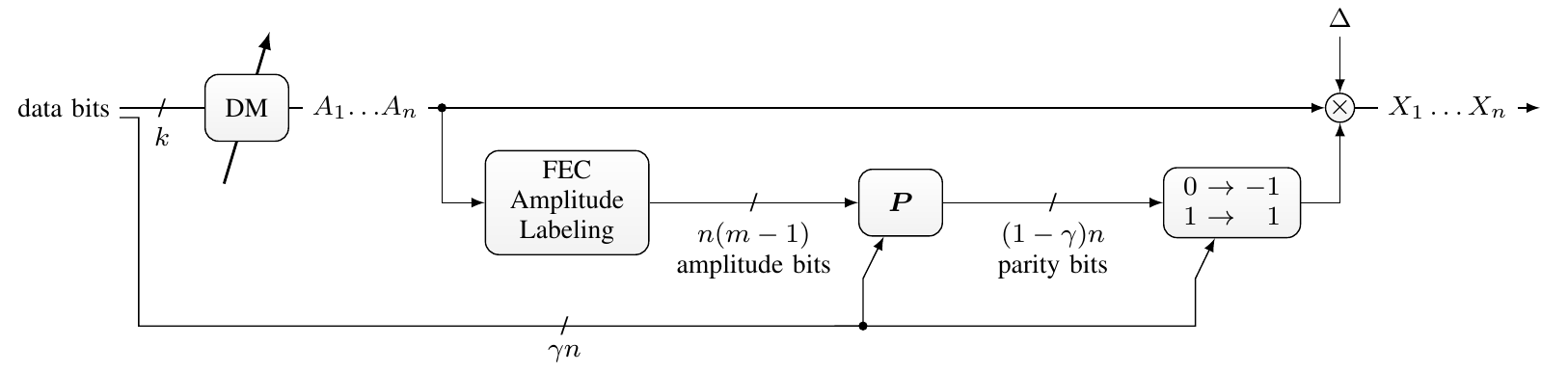}
\caption{The architecture of the \ac{PAS} scheme. See Fig.~\ref{fig:system_model} for a system view and \cite[Sec.~IV.]{bocherer2015bandwidth} for a detailed description.}
\label{fig:pas internals}
\end{figure*}

\subsection{Distribution Matching (DM)}
\label{sec:dm}

\ac{DM}s~\cite{bocherer2011matching,schulte2016constant} transform a sequence of uniformly distributed input bits into an output sequence of symbols from an alphabet $\seta$ with a desired distribution. A fixed-to-fixed length \ac{DM} maps $k$ input bits $d^k$ to $n$ output symbols $a^n=\dm(d^k)$. The mapping $\dm$ is invertible, i.e., $d^k$ can be recovered from $a^n$ by applying the inverse mapping $\dm^{-1}$. Fixed-to-fixed length \ac{DM}s can be implemented by the \ac{CCDM}~\cite{schulte2016constant}, for binary output alphabets see also \cite{ramabadran_coding_1990}. A \ac{DM} is characterized by the following parameters.
\begin{itemize}
\item The rate is
\begin{align}
\rdm=\frac{k}{n}\quad\left[\frac{\text{bits}}{\text{output symbol}}\right].
\end{align}
\item The output distribution is
\begin{align}
P_A(a)=\frac{\sum_{d^k\in\{0,1\}^k}P_{\dm(d^k)}(a)}{2^k},\quad a\in\seta
\end{align}
where $P_{a^n}$ is the empirical distribution of the sequence $a^n$, i.e.,
\begin{align}
P_{a^n}(a)=\frac{|\{i\colon a_i=a\}|}{n},\quad a\in\seta.
\end{align}
\item The rate loss is the difference of the \ac{DM} rate and the entropy rate of a \ac{DMS} $P_A$, i.e.,
\begin{align}
\rloss=\entop(A)-\frac{k}{n}.\label{eq:rate loss dm}
\end{align}
\end{itemize}
By \cite[Sec.~III.B]{schulte2016constant}, the rate loss of \ac{CCDM} vanishes for large output lengths $n$. In this work, we are interested in \ac{DM}s with relatively short output lengths and we therefore need to account for the rate loss in our system design.

\subsection{Amplitude Shift Keying Modulation}

We consider $2^m$-\ac{ASK} constellations
\begin{align}
\mathcal{X} = \left\{\pm1,\pm3,\ldots,\pm (2^m-1)\right\}
\end{align}
with amplitude alphabet
\begin{align}
\mathcal{A} = \left\{1,3,\dotsc,2^m-1\right\}.
\end{align}
We use label functions $\beta\colon\setx\to\{0,1\}^m$ and corresponding bit mappers $\chi\colon\{0,1\}^m\to\setx$. For all labels $\vecB=B_1\dotsc B_m$ in this work, the first bit $B_1$ labels the sign $S$ according to
\begin{align}
B_1=\begin{cases}0,&S=1\\
1,&S=-1.
\end{cases}
\end{align}
Consequently, $B_2\dotsc B_m$ label the amplitudes and each label function $\beta$ implies an amplitude label function $\beta_A$ and each bit-mapper $\chi$ implies an amplitude bit-mapper $\chi_A$. Two labels are of special interest, namely the \ac{BRGC}~\cite{gray1953pulse} and the \ac{NBBC}~\cite[Sec.~VI.C]{bocherer2015bandwidth} where the amplitude label is a \ac{NBC}. The two labels are illustrated for 8-ASK in Table~\ref{tab:labels}.
\newcommand{\tb}[1]{\textbf{#1}}
\begin{table}
\caption{Two labels for 8-ASK. The amplitude label of NBBC is NBC and the amplitude label of BRGC is also BRGC.}
\label{tab:labels}
\begin{tabular}{rrrrrrrrr}
\toprule
&-7&-5&-3&-1&1&3&5&7\\\midrule
BRGC&000&001&011&010&1\tb{10}&1\tb{11}&1\tb{01}&1\tb{00}\\
NBBC&000&001&010&011&1\tb{11}&1\tb{10}&1\tb{01}&1\tb{00}\\\bottomrule
\end{tabular}
\end{table}

\subsection{\ac{PAS} Transmitter}
\label{sec:pas}

The \ac{PAS} architecture implements probabilistically shaped ASK modulation. The \ac{PAS} transmitter is displayed in Fig.~\ref{fig:pas internals} and works as follows (for a more detailed description, see \cite[Sec.~IV.]{bocherer2015bandwidth}). A \ac{DM} maps $k$ data bits to $n$ amplitudes $A^n$, which are represented by $n(m-1)$ amplitude bits. The amplitude bits and $\gamma n$ additional data bits are multiplied with the parity generating part $\matp$ of a systematic generator matrix $[\mati|\matp]$ to generate $(1-\gamma)n$ redundancy bits. The redundancy bits and the additional data bits are mapped to $n$ signs $S^n$, which are multiplied symbolwise with the amplitudes $A^n$. The FEC code instantiated by $\matp$ has rate
\begin{align}
c=\frac{n(m-1)+\gamma n}{mn}=\frac{m-1+\gamma}{m}\label{eq:fec c}
\end{align}
and the fraction of signs used for data bits is
\begin{align}
\gamma = 1-(1-c) m.\label{eq:gamma}
\end{align}
\ac{PAS} requires $0\leq\gamma\leq 1$. The transmission rate of \ac{PAS} is the number of data bits per \ac{ASK} symbol given by
\begin{align}
\rtx=\frac{k}{n}+\gamma.
\end{align}
\subsection{Channel Model}
The generated signal points $A_i\cdot S_i$ are multiplied by the constellation scaling $\Delta$ and transmitted over an \ac{AWGN} channel. We define $X_i=\Delta \cdot A_i \cdot S_i$. At a generic time instance, the channel model is
\begin{align}
Y=X+Z
\end{align}
where $Z$ is zero mean Gaussian noise with unit variance. The \ac{SNR} is
\begin{align}
\snr=\expop[X^2].
\end{align}
\subsection{\ac{PAS} Achievable Rate}
\label{sec:pas_achievable_rate}
We consider a \ac{PAS} receiver with a bit-metric decoder. The \ac{PAS} transmitter defines the label $\vecB^\text{fec}=\beta^\text{fec}(X)$ where $B_1^\text{fec}=B_1$ is the sign label and where $B_2^\text{fec}\dotsc B_m^\text{fec}$ is the amplitude label. We assume a uniform sign distribution, i.e.,
\begin{align}
P_{B_1}(0)=P_{B_1}(1)=P_S(-1)=P_S(1)=\frac{1}{2}.
\end{align}
We refer to \cite[Sec.~IV.A]{bocherer2015bandwidth} for a justification of this assumption. A binary demapper calculates the soft-informations
\begin{align}
    L_j=\log\frac{P_{B^\text{fec}_j}(0)}{P_{B^\text{fec}_j}(1)}+\log\frac{p_{Y|B^\text{fec}_j}(y|0)}{p_{Y|B^\text{fec}_j}(y|1)},\quad j=1,2,\dotsc,m
\end{align}
which are passed to a binary decoder. By \cite{bocherer2014achievable}, an achievable rate for a bit-metric decoder is
\begin{align}
    \rbmd = \left[\entop(X)-\sum_{j=1}^m \entop(B^\text{fec}_j|Y)\right]^+\label{eq:Rbmd_general}
\end{align}
where $[\cdot]^+ = \max(0,\cdot)$.
For \ac{PAS}, $\rbmd$ must be evaluated using $P_X=P_AP_S$. The rate $\rbmd$ is an achievable rate for the \ac{PAS} receiver with BMD if
\begin{align}
\entop(A)+\gamma = \rbmd.\label{eq:pas achievable rate}
\end{align}
We will also need the relations between $\rbmd$, $P_X$, and $\snr$, namely
\begin{align}
R&=\rbmd(P_X,\snr)\\
\snr&=\rbmd^{-1}(P_X,R).
\end{align}

\begin{figure}
    \centering
    \footnotesize
    \tikzsetnextfilename{8ask_rates}
    \includegraphics{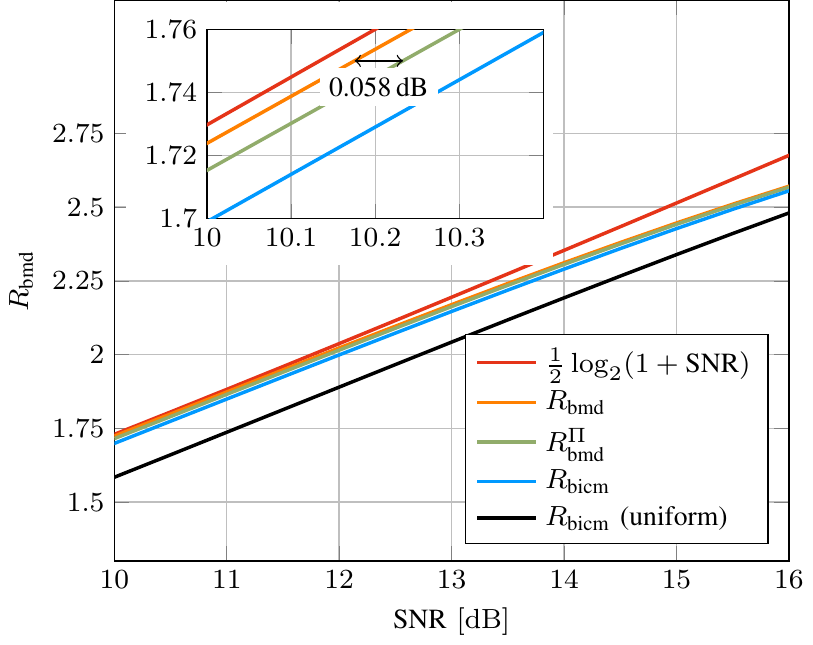}
    \caption{Achievable rates for 8-ASK.}
    \label{fig:8ask_rates}
\end{figure}

\section{Product Distribution Matching}
\label{sec:pdm}

\subsection{NBC Product Distributions}

Suppose for some amplitude label $B_2^\text{dm}\dotsb B_m^\text{dm}$ and the corresponding signal point label $\vecB^\text{dm}=B_1B_2^\text{dm}\dotsb B_m^\text{dm}$ we have
\begin{align}
P_{\vecB^\text{dm}}=\prod_{j=1}^m P_{B^\text{dm}_j}\label{eq:product constraint}
\end{align}
where $P_{B_1^\text{dm}}=P_{B_1}$. In particular, the amplitude distribution $P_A$ is such that the bits of the label $\vecB^\text{dm}$ are statistically independent. We can construct a distribution \eqref{eq:product constraint} by choosing an amplitude label and binary distributions $P_{B_j^\text{dm}}$, $j=2,\dotsc, m$. Note that the generated distribution depends both on the label function and the binary distributions. An achievable rate is
\begin{align}
    \prbmd = \left[\sum_{j=1}^m\entop(B^\text{dm}_j)-\sum_{j=1}^m \entop(B^\text{fec}_j|Y)\right]^+\label{eq:Rbmd_product_entropies}.
\end{align}
Note that the label $\vecB^\text{dm}$ is not required to be the same as the label $\vecB^\text{fec}$ that is used by the FEC encoder and decoder.
We choose the \ac{NBC} for the amplitude label $B_2^\text{dm},\dotsc,B_m^\text{dm}$, the \ac{BRGC} for the \ac{FEC} label $\vecB^\text{fec}$
and we optimize \eqref{eq:Rbmd_product_entropies} over the binary distributions $P_{B_j^{\text{dm}}}$, $j=2,\dots,m$ (recall that the sign distribution $P_{B_1}$ is uniform) and the constellation scaling $\Delta$. In Fig.~\ref{fig:8ask_rates},
we display the resulting achievable rate for 8-ASK. We observe that the product constraint \eqref{eq:product constraint} leads to virtually no performance loss.

\begin{remark}
The information-theoretic work \cite{ifabregas2010bit} considered only the case when $\vecB^\textnormal{dm} = \vecB^\textnormal{fec}$, in which case \eqref{eq:Rbmd_product_entropies} becomes
\begin{align}
    \rbicm = \sum_{j=1}^m \miop(B_j^\textnormal{fec};Y)\label{eq:Rbmd_deg}
\end{align}
which is the so-called BICM capacity. As shown in Fig.~\ref{fig:8ask_rates}, $\rbicm$ is less power efficient than $\prbmd$, although the difference is small.
\end{remark}

\subsection{PDM}
\ac{PDM} can efficiently generate the product distributions introduced in the previous subsection. The \ac{PDM} is displayed in Fig.~\ref{fig:pdm}. $k$ binary data bits are demultiplexed into $m-1$ parallel blocks of lengths $k_2$ to $k_m$. The $m-1$ parallel binary \ac{DM}s output $m-1$ shaped binary sequences of length $n$. A bit mapper  $\chi_A^\text{dm}$ recombines the $m-1$ sequences and outputs one shaped amplitude sequence of length $n$.

\subsection{PDM Rate Loss}

The rate and the output distribution of the $j$th DM is $k_j/n$ and $P_{B_j^\text{dm}}$, respectively. The total rate of the PDM is
\begin{align}
\frac{k}{n}=\frac{k_2+\dotsb+k_m}{n}
\end{align}
and the total rate loss of the PDM is the sum of the individual rate losses, i.e.,
\begin{align}
   \rloss= \sum_{j=2}^m \left[\entop(B_j^\text{dm})-\frac{k_j}{n}\right].\label{eq:rate loss pdm}
\end{align}

\subsection{PDM for the AWGN Channel}
\label{sec:pdm_for_awgn}

For the AWGN channel, we use the \ac{NBC} for the bit-mapper $\chi_A^\text{dm}$ and we choose binary \ac{DM} distributions that minimize the overall power. Ignoring the rate loss for now, the optimization problem is
\begin{align}
\begin{split}
\minimize_{P_{B_2},\dotsc,P_{B_m}}\quad&\expop[X^2]\\
\st\quad&\sum_{j=2}^m\entop(B_j)=\rdm\\
&X=\chi^\text{nbbc}(\vecB).
\end{split}\label{eq:opt_entropy_constrained_dm}
\end{align}
To account for the rate loss, we replace the sum-entropy constraint in \eqref{eq:opt_entropy_constrained_dm} by a sum-rate constraint, where the $j$th rate $k_j/n$ is the rate required to implement the \ac{DM} output distribution $P_{B_j}$. Altogether, we choose the component \ac{DM}s via
\begin{align}
\begin{split}
\minimize_{P_{B_2},\dotsc,P_{B_m}}\quad&\expop[X^2]\\
\st\quad&\sum_{j=2}^m\frac{k_j}{n}=\rdm\\
&X=\chi^\text{nbbc}(\vecB).
\end{split}\label{eq:opt_entropy_constrained_real}
\end{align}

\subsection{Simulation Results}
\label{sec:sim_results}
We numerically compare different \ac{DM} implementations by using 64-ASK and a target \ac{SE} of $R_\text{t} = \SI{4.5}{\bpcu}$. We employ a 32-ary \ac{DM} as a reference as suggested in~\cite[Sec. V]{bocherer2015bandwidth}. The performance of this system is compared to a \ac{PDM} setup with 1 $(B^\text{dm}_2)$, 2 $(B^\text{dm}_2, B^\text{dm}_3)$, 3 $(B^\text{dm}_2, B^\text{dm}_3, B^\text{dm}_4)$, 4 $(B^\text{dm}_2, B^\text{dm}_3, B^\text{dm}_4, B^\text{dm}_5)$ and 5 $(B^\text{dm}_2, B^\text{dm}_3, B^\text{dm}_4, B^\text{dm}_5, B^\text{dm}_6)$ individually shaped bit-levels and corresponding binary \acp{DM}. The product distribution has been obtained by following the approach of Sec.~\ref{sec:pdm_for_awgn}, while imposing a uniform distribution on the unshaped bit-levels.

\begin{figure}[t]
\centering
\footnotesize
\tikzsetnextfilename{dm_loss}
\includegraphics{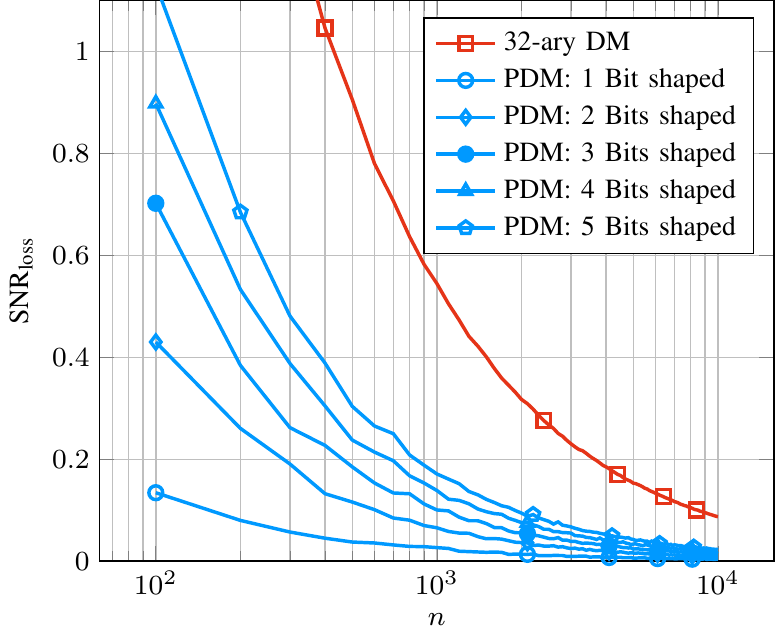}
\caption{Rate loss comparison for 64-ASK and $\text{SE} = \SI{4.5}{\bpcu}$.}
\label{fig:dm_loss}
\end{figure}

We first consider the results of Fig.~\ref{fig:dm_loss} which illustrates the finite length loss of all considered configurations. The \ac{DM} rate loss \eqref{eq:rate loss dm} and the \ac{PDM} rate loss \eqref{eq:rate loss pdm} is converted to an ``SNR loss'' by
\begin{equation}
\text{SNR}_{\text{loss}} = 10\log_{10}\left(\frac{\rbmd^{-1}(P_X,\rtx+\rloss)}{\rbmd^{-1}(P_X,\rtx)}\right).\label{eq:snr_loss}
\end{equation}
As a rule of thumb, the following expression may be useful as a rough estimate:
\begin{align}
\text{SNR}_\text{loss,awgn} &= 10\log_{10}\left(\frac{2^{2(\rtx+\rloss)}-1}{2^{2\rtx}-1}\right)\nonumber\\
&\approx \rloss \cdot 20\log_{10}2\approx\rloss\cdot\SI{6}{dB}.\label{eq:snr_loss_awgn}
\end{align}
We observe that the \acp{PDM} have an aggregated rate loss that  is significantly lower than the rate loss of the 32-ary \ac{DM}. The resulting performance is comparable only for output lengths of more than \num{e4}
symbols.

To further illustrate the flexibility of the transmitter design, we consider a coded scenario with a rate 9/10 \ac{LDPC} block code from the DVB-S2 standard~\cite{etsi2009dvb} of block length \num{64800} bits and a corresponding \ac{DM} output length of \num{10800} symbols. This choice allows for a fair comparison, as both the
parallel binary \acp{DM} and the 32-ary \ac{DM} have a similar performance. Fifty iterations are used for the \ac{BP} decoding.

\begin{figure}[t]
\centering
\footnotesize
\tikzsetnextfilename{coded_results}
\includegraphics{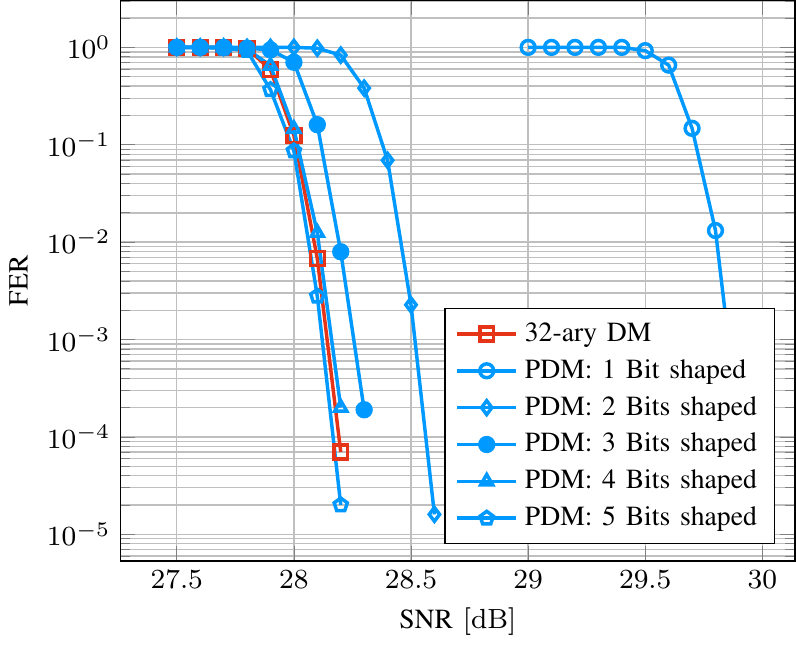}
\caption{Performance comparison of the proposed PDM for 64-ASK and a target \ac{SE} of \SI{4.5}{\bpcu} and different number of shaped bits.}
\label{fig:coded_results1}
\end{figure}

As shown in Fig.~\ref{fig:coded_results1}, a \ac{PDM} with 3 shaped bit-levels achieves a similar performance as the 32-ary \ac{DM}. If only 2 bit-levels are shaped, the loss in energy efficiency is \SI{0.4}{dB} at a target \ac{FER} of \num{e-3}. Table~\ref{tab:coding_required_snr} illustrates that these observations are reflected by the asymptotic achievable rates of Sec.~\ref{sec:pas_achievable_rate}, which were evaluated for the
corresponding optimized distributions. While the required \acp{SNR}
to achieve an \ac{SE} of \SI{4.5}{\bpcu} are close for 3, 4 and 5 shaped bit-levels, larger gaps can be observed for 1 or 2 shaped bit-levels.

\begin{table}
 \footnotesize
 \centering
 \caption{Required \text{SNRs} for different \ac{DM} configurations and a target \ac{SE} of \SI{4.5}{\bpcu}. (Capacity: \SI{27.08}{dB})}
 \label{tab:coding_required_snr}
 \begin{tabular}{lc}
 \toprule
 DM configuration & Required \text{SNR} [\si{dB}]\\
 \midrule
 32-ary DM & \num{27.13}\\
 PDM 1 Bit shaped & \num{28.29}\\
 PDM 2 Bits shaped & \num{27.48}\\
 PDM 3 Bits shaped & \num{27.35}\\
 PDM 4 Bits shaped & \num{27.32}\\
 PDM 5 Bits shaped & \num{27.31}\\
 \bottomrule
 \end{tabular}
\end{table}

\section{Probabilistic Shaping for Parallel Channels}
\label{sec:parallel_channel}

\begin{figure*}
	    \centering
	    \tikzsetnextfilename{parallelPAS}
    \includegraphics{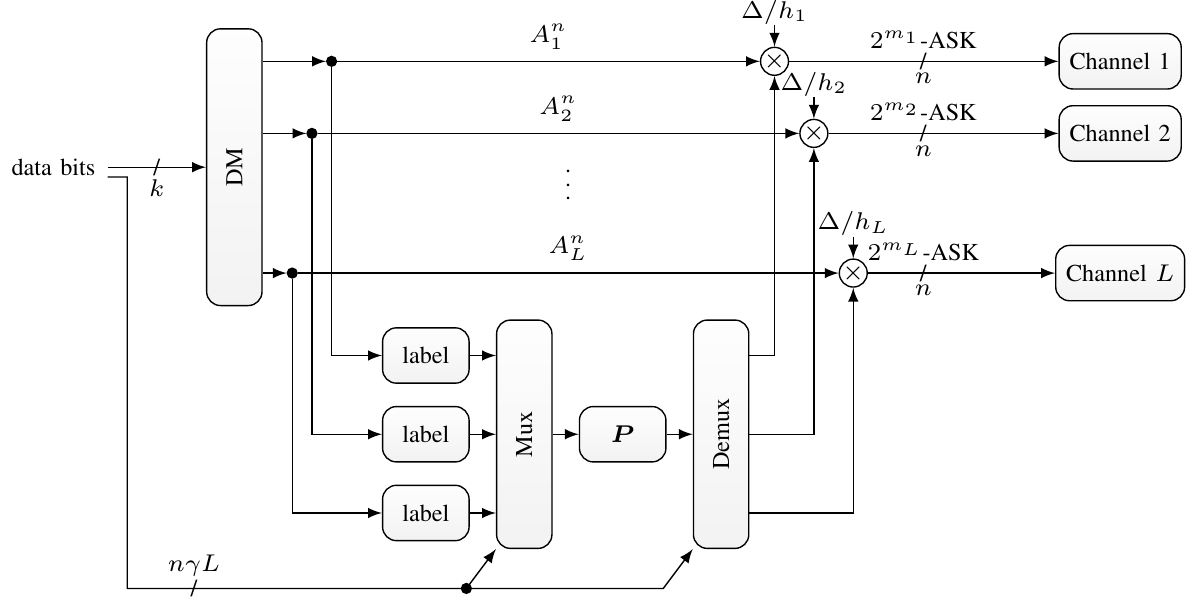}
    \caption{Illustration of \ac{PAS} for parallel channels. Each of the $L$ channels has an individual constellation size $2^{m_\ell}$, $\ell=1,2,\dotsc,L$. PAS for parallel channels extends the PAS for single channels shown in Fig.~\ref{fig:pas internals}. The constellation scalings $\Delta/h_\ell$ are explained in Sec.~\ref{subsec:pas waterfilling}. The DM device can be implemented by $L$ individual DMs, as illustrated in Fig.~\ref{fig:parallel_ref_dm}, or by extended PDM illustrated in Fig.~\ref{fig:extendedPDM}. Extended PDM can be much more power efficient, see Fig.~\ref{fig:coded_results_parallel} for an example.}
    \label{fig:parallelPAS}
    \tikzsetnextfilename{naive_DM_parallel_CH}
    \includegraphics{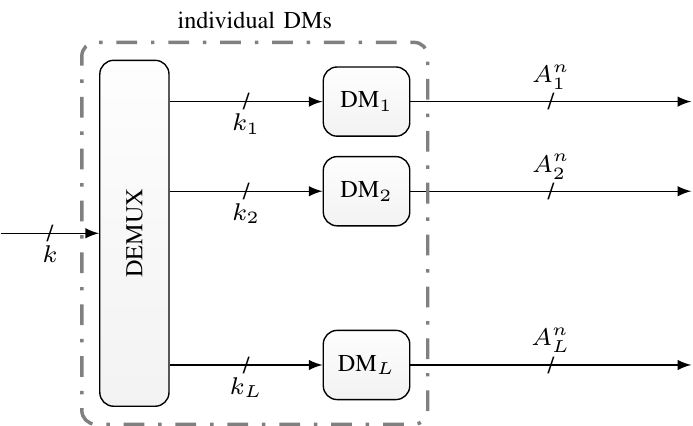}
	\caption{Individual DM implementation for PAS for parallel channels shown in Fig.~\ref{fig:parallelPAS}. The $k$ data bits are demultiplexed and fed to $L$ individual DMs.}
	\label{fig:parallel_ref_dm}
	\tikzsetnextfilename{Layered_DM}
	\includegraphics{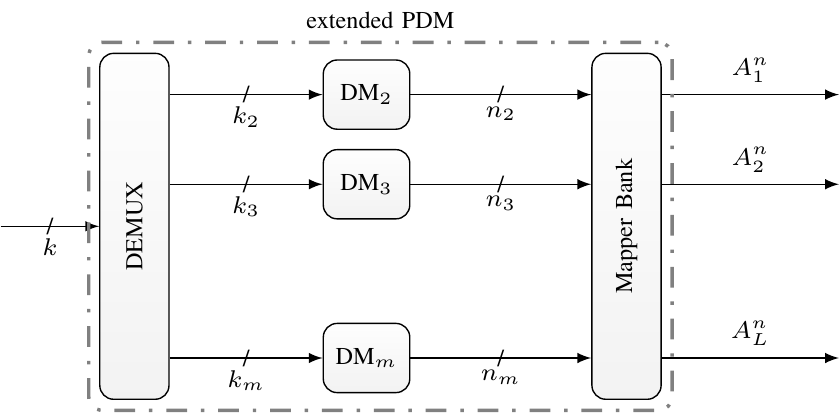}
	\caption{Extended \ac{PDM} implementing the \ac{DM} of PAS for parallel channels shown in Fig.~\ref{fig:parallelPAS}. The extended PDM transforms $k$ data bits into $L$ amplitude sequences of length $n$. Internally, the extended \ac{PDM} uses $m-1$ binary component \ac{DM}s, where $2^m$-ASK is the largest supported constellation. The output lengths $n_2,\dotsc,n_m$ of the component \ac{DM}s are given by \eqref{eq:component dm nj}. The input lengths fulfill $\sum_{j=2}^m k_j=k$ and reflect the component \ac{DM} rates $k_j/n_j$.}
	\label{fig:extendedPDM}
\end{figure*}

\subsection{System Model}
We consider $L$ parallel channels with the I/O relation
\begin{align}
    Y_\ell = h_\ell X_\ell+Z_\ell,\quad \ell=1,2,\dotsc,L.
\end{align}
The noise terms $Z_\ell$ are zero mean Gaussian with unit variance. The $h_\ell$ model the channel gains and we assume that both the receiver and transmitter have full channel state information, i.e., they both know the channel gains $h_\ell$ and the noise variance. We consider coding over $n$ channel uses of each channel, which results in total in $L\cdot n$ channel uses. This choice is for clarity of exposition; the scheme can easily be generalized.
\subsection{Waterfilling \cite[Sec.~5.4.6]{tse2005fundamentals}}
\label{sec:waterfilling}
The transmitter has an average power budget $P$, i.e., the inputs are subject to the sum-power constraint
\begin{align}
    \frac{1}{L}\sum_{\ell=1}^L\expop[X_\ell^2]\leq P.
\end{align}
The average \ac{SE}
\begin{align}
    \frac{1}{L}\sum_{\ell=1}^L\frac{1}{2}\log_2(1+h_\ell^2P_\ell)
\end{align}
is achievable with the channel inputs $X_\ell$ being independent zero mean Gaussian with variance $P_\ell$. The average \ac{SE} is maximized by waterfilling, i.e.,
\begin{align}
    P_\ell^*=\left[\frac{1}{\lambda}-\frac{1}{h_\ell^2}\right]^+,\quad\lambda\colon \frac{1}{L}\sum_{\ell=1}^L P_\ell^*=P.
\end{align}
Suppose that $P_\ell^*$ is positive. The SE allocated to channel $\ell$ is then
\begin{align}
    C_\ell=\frac{1}{2}\log_2\frac{h_\ell^2}{\lambda}.
\end{align}
Based on $C_\ell$, we choose the constellation size $2^{m_\ell}$ so that
\begin{align}
    m_\ell\approx C_\ell+1\label{eq:rule_cstll_size}
\end{align}
to avoid reduced \ac{SE} because of too small constellation sizes. Let $m=\max_\ell m_\ell$ denote the maximum constellation size.

\subsection{PAS for Parallel Channels}
\label{sec:pas_parallel_channels}

\ac{PAS} can easily be combined with parallel channels. This is illustrated in Fig.~\ref{fig:parallelPAS}. A \ac{DM} device transforms data bits into a sequence of amplitudes for each channel, which are then combined with sign bits originating from a common encoding device. In its simplest form, this \ac{DM} device consists of individual \acp{DM}, each with its output alphabet size matched to the corresponding constellation size, see Fig.~\ref{fig:parallel_ref_dm}.

\subsection{\ac{PDM} for Parallel Channels}
\label{sec:pdm_parallel_channels}

\begin{figure*}[t]
\centering
\footnotesize
\tikzsetnextfilename{bitchanneldistributions}
\includegraphics{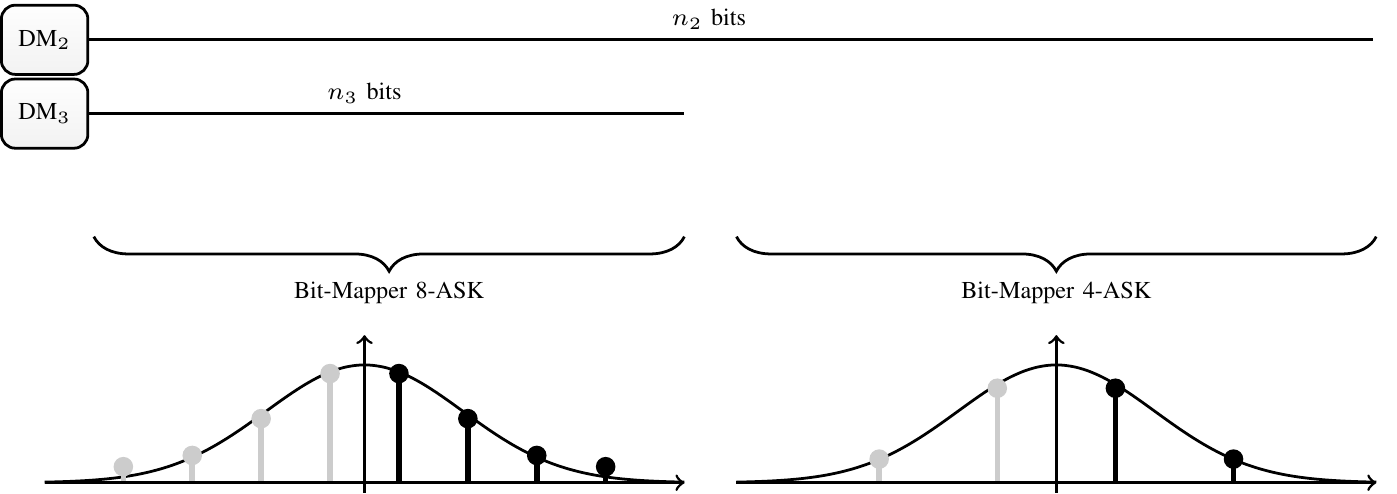}
\caption{Simultaneously generating two Gaussian-like amplitude distributions for 4-ASK and 8-ASK by reusing the \ac{DM} of bit-level 2.}
\label{fig:many gaussian}
\end{figure*}

The PDM suggests an alternative way to generate $L$ amplitude sequences for distinct constellation sizes. For example, suppose we have $L=2$ different channels and need a length $n$ amplitude sequence for 4-ASK and a length $n$ sequence for 8-ASK. The PDM needs one binary \ac{DM} for 4-ASK and two binary \ac{DM}s for 8-ASK. As illustrated in Fig.~\ref{fig:many gaussian}, the idea is now to use for the first amplitude bit-level $B_2$ of 4-ASK and 8-ASK a single binary \ac{DM} with output length $n_2=2\cdot n$ and to generate the second amplitude bit-level $B_3$ for 8-ASK by a second binary \ac{DM} with output length $n_3=n$. The potential benefit of this approach is twofold: first, using \ac{PDM} should reduce the rate loss, and second, replacing two \ac{DM}s of lengths $n$ by one single \ac{DM} of length $2\cdot n$ should reduce the rate loss even further. Fig.~\ref{fig:extendedPDM} shows this extended PDM scheme. It provides the same interface to PAS as the naive approach that uses $L$ individual \ac{DM}s.

Simultaneously using one \ac{DM} on more than one constellation size imposes restrictions on the distribution families that can be generated by extended PDM. We next argue how extended PDM can be used to generate families of Gaussian-like distributions. The maximum costellation size is $2^m$ and we choose the $m-1$ \ac{DM} output distributions so that an \ac{NBBC} mapper generates a Gaussian-like distribution. By grouping $2^j$ neighbouring signal points together, the distribution of theses signal point groups is still Gaussian-like, and it is given by the product distribution generated by the first $m-1-j$ \ac{DM}s. This suggests that by using only $m-1$ \ac{DM}s, we can simultaneously generate Gaussian-like distributions on $4,8,\dotsc,2^m$-ASK constellations. An example is shown in Fig.~\ref{fig:many gaussian}.


\subsection{Parametrization}
\label{sec:pdm_parametrization}

We next state the parameters of the \ac{FEC} code and the \ac{PDM} so that the parallel \ac{PAS} operates at a specific \ac{SE}. For the considered case where we use each of the $L$ channels $n$ times, the block length of the binary FEC code is
\begin{align}
n_\text{code}=\sum_{\ell=1}^L m_\ell\cdot n
\end{align}
and formulas \eqref{eq:fec c} and \eqref{eq:gamma} generalize to
\begin{align}
c&=\frac{\sum_{\ell=1}^L(m_\ell-1+\gamma)}{\sum_{\ell=1}^L m_\ell}\\
\gamma&=1-(1-c)\frac{1}{L}\sum_{\ell=1}^L m_\ell.
\end{align}
The \ac{DM} output lengths are given by
\begin{align}
n_j=\sum_{\ell=1}^L\mathds{1}(m_\ell\geq j)\cdot n,\quad j=2,3,\dotsc,m\label{eq:component dm nj}
\end{align}
and the corresponding \ac{DM} input lengths are $k_2,k_3,\dotsc,k_m$. The average \ac{SE} of the overall system is now
\begin{align}
\rtx&=\frac{\sum_{j=2}^m k_j}{L\cdot n}+\gamma\label{eq:ePDM rate}\\
&=\frac{1}{L\cdot n}\Bigl[\sum_{j=2}^m\entop(B_j^\text{dm})n_j\Bigr]+\gamma-\rloss.\label{eq:ePDM entropy}
\end{align}

\subsection{Waterfilling for PAS}\label{subsec:pas waterfilling}
For the $L$ parallel channels, suppose we have chosen the constellation sizes $2^{m_\ell}$, $\ell=1,\dotsc,L$ and suppose further we have chosen the code rate $c$ and thereby the fraction $\gamma$ of signs used for data bits. To achieve the target rate $\rtx$, the rate assigned to the amplitudes is thus $\rdm = \rtx-\gamma$, which results in the following constraint for the amplitude distributions (ignoring the rate loss):
\begin{align}
\frac{1}{L}\sum_{\ell=1}^L\entop(A_\ell)=\rdm.
\end{align}
Recall that the inputs $X_\ell$ are given by $\Delta_\ell A_\ell S_\ell$ where
\begin{align}
A_\ell S_\ell\in\{\pm 1,\pm 3,\dotsc,\pm (2^{m_\ell}-1)\}.
\end{align}
The average power on the $\ell$th channel is $\expop[(\Delta_\ell A_\ell)^2]$ and depends on the distribution $P_{A_\ell}$ and the constellation scaling $\Delta_\ell$. We use the following strategy: to ensure a similar detection reliability on each channel, independent of the chosen amplitude distributions, we choose
\begin{align}
    \Delta_\ell=\frac{\Delta}{h_\ell}.
\end{align}
In this way, two neighbouring constellation points have the distance $2\Delta$ on all channels. The average power on each channel is $\frac{\Delta^2}{h_\ell^2}\expop[A_\ell^2]\propto \frac{1}{h_\ell^2}\expop[A_\ell^2]$. Next, we calculate the amplitude distributions by
\begin{align}
\minimize_{P_{A_1},\dotsc,P_{A_L}}\quad&\sum_{\ell=1}^L \frac{1}{h_\ell^2}\expop[A_\ell^2]\\
\st\quad&\frac{1}{L}\sum_{\ell=1}^L\entop(A_\ell)=\rdm.
\end{align}
To account for rate loss, the sum-entropy constraint is replaced by a \ac{DM} sum-rate constraint. For extended \ac{PDM}, the sum-entropy and sum-rate expressions from \eqref{eq:ePDM entropy} and \eqref{eq:ePDM rate} are used, respectively.

\subsection{Simulation Results}

\begin{figure}
\centering
\footnotesize
\tikzsetnextfilename{coded_results_parallel}
\includegraphics{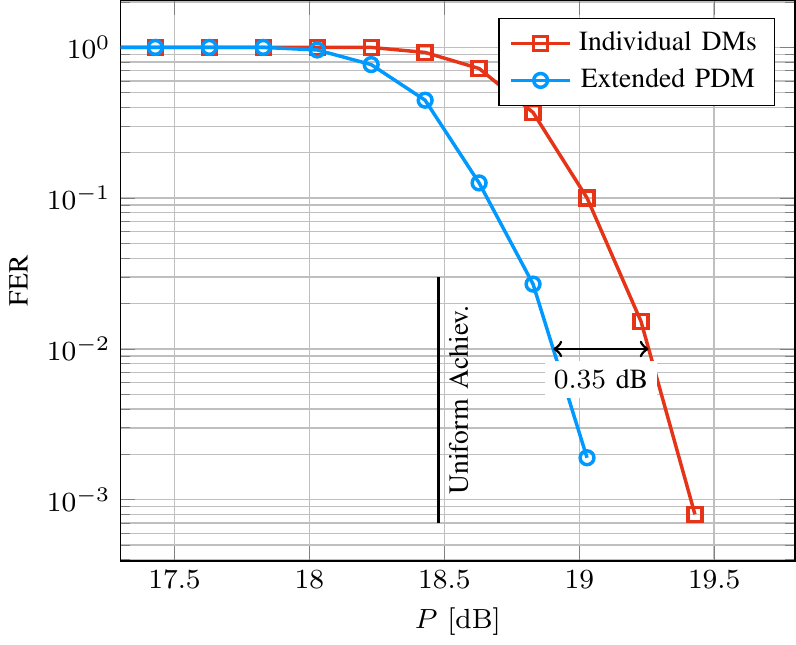}
\caption{Coded performance comparison of individual DMs and extended PDM (LDPC code with block length \num{5184} bits) for parallel channels.}
\label{fig:coded_results_parallel}
\end{figure}

\begin{figure}
\centering
\footnotesize
\tikzsetnextfilename{rates_parallel_asymptotic}
\includegraphics{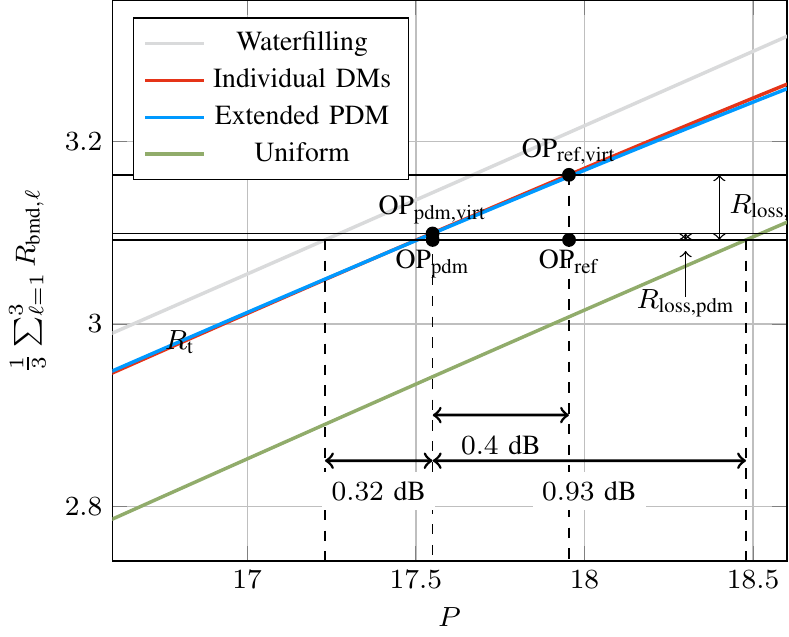}
\caption{Asymptotic analysis of the incurred rate loss.}
\label{fig:rates_parallel_asymptotic}
\end{figure}

To evaluate the performance of parallel \ac{PAS} and extended \ac{PDM}, we employ the following example of 3 parallel channels, given as
\begin{align*}
    Y_1 &= 2.0 \cdot X_1 + Z_1\\
    Y_2 &= 1.3 \cdot X_2 + Z_2\\
    Y_3 &= 0.6 \cdot X_3 + Z_3
\end{align*}
and an average power constraint of $P = \SI{17.23}{dB}$. Performing the waterfilling as shown in Sec.~\ref{sec:waterfilling}, we arrive at the following rate assignment:
\begin{align*}
    C_1 &= \num{3.87}\\
    C_2 &= \num{3.25}\\
    C_3 &= \num{2.14}.
\end{align*}
Consequently, we target an \ac{SE} of $R_\text{t} = \frac{1}{3} \sum_{\ell=1}^3 C_\ell = \SI{3.09}{\bpcu}$ and select constellation sizes of $2^{m_1} = 32$, $2^{m_2} = 16$ and $2^{m_3} = 8$ points following~\eqref{eq:rule_cstll_size}. We use each of the three channels $n=\num{432}$ times.

As a reference, we choose an architecture with individual 16-ary, 8-ary and 4-ary \acp{DM}.
For the \ac{PDM} setup, we employ four parallel binary \acp{DM}. Their respective output lengths and distributions are summarized in Table~\ref{tab:parallel_channels_dm_lengths}. We have a maximum constellation size of 32-ASK, i.e., four bits can be shaped. Bit-levels 2 and 3 are shared by all three constellations, whereas bit-level 4 is used only by 16-ASK and 32-ASK. Bit-level 5 appears in 32-ASK only.
\begin{table}[t]
\caption{Properties of the extended PDM setup}
\label{tab:parallel_channels_dm_lengths}
\centering
\begin{tabular}{llll}
\toprule
$\text{DM}_j$ & $n_j$ & $P_{B_{j}^\text{dm}}(0)$ & $\entop(B_j^\text{dm})$\\
\midrule
2 & 1296 & \num{0.2522} & \num{0.8148}\\
3 & 1296 & \num{0.3940} & \num{0.9674}\\
4 & 864 & \num{0.4474} & \num{0.9920}\\
5 & 432 & \num{0.4674} & \num{0.9969}\\
\bottomrule
\end{tabular}
\end{table}
The distributions of the individual DMS and the binary distribution of the extended PDM have been chosen following Sec.~\ref{subsec:pas waterfilling}.

In the following, we use a block length \num{5184} \ac{LDPC} code of rate $c = 5/6$ from the G.hn standard~\cite{itu_ghn}. As before, 50 \ac{BP} iterations are performed.

Observe in Fig.~\ref{fig:coded_results_parallel} that the PDM setup improves over the reference strategy at a \ac{FER} of \num{e-2} by \SI{0.35}{dB}. This is mainly because of the decreased rate loss as shown in the asymptotic achievability plot of Fig.~\ref{fig:rates_parallel_asymptotic}. We plot the average achievable rate over all parallel channels vs. the average sum power for both schemes and their specific input distributions. The power assignment is optimized via mercury/waterfilling. We also plot three horizontal lines at \SI{3.09}{\bpcu}, \SI{3.099}{\bpcu}
and \SI{3.16}{\bpcu}, which denote $\rtx$, $\rtx+R_{\text{loss,pdm}}$ and $\rtx+R_{\text{loss,ref}}$, respectively. The crossing of the last two horizontal lines with their respective achievability curves are labeled as $\text{OP}_\text{pdm,virt}$ and $\text{OP}_\text{ref,virt}$. They indicate virtual operating points that would be achievable with the currently used input distributions. Because of the rate loss, the actual operating points are given by the orthogonal projections of these points on the actual \ac{SE} curve, however. Their difference in SNR of \SI{0.4}{dB} accurately predicts the gap of \SI{0.35}{dB} that we observe in the coded result in Fig.~\ref{fig:coded_results_parallel}. Compared to uniform distributions, the asymptotic gain (accounting for the rate loss) is \SI{0.93}{dB}. The gap to the waterfilling solution is \SI{0.32}{dB}.

\section{Conclusion}
\label{sec:conclusion}

We proposed product distribution matching (PDM), an architecture that uses binary \ac{DM}s in parallel. This parallelization enables high-throughput implementations of \ac{DM}s. The binary component \ac{DM}s of \ac{PDM} reduce complexity. We have shown that \ac{PDM} performs as well as higher-order \ac{DM}s for long block lengths and that \ac{PDM} can perform much better than higher-order \ac{DM}s for short block lengths. We have proposed extended \ac{PDM}, which enables PAS to operate close to the waterfilling limit of multi-carrier transmission schemes such as OFDM.

\section*{Acknowledgment}
The authors would like to thank Gerhard Kramer for fruitful discussions and comments on drafts of this manuscript.


\end{document}